\documentclass[journal]{IEEEtran}
\usepackage{amsthm}
\usepackage{latexsym}
\usepackage{amssymb}
\usepackage[dvips]{graphics}
\usepackage{fancyhdr}
\usepackage{amsmath}
\usepackage{color}
\usepackage{epsfig}
\usepackage{cite}
\usepackage{cleveref}

\theoremstyle{plain}
\newtheorem{theorem}{Theorem}
\newtheorem{lemma}{Lemma}

\newtheorem{corollary}{Corollary}
\theoremstyle{definition}

\theoremstyle{remark}

\newcommand{\off}[1]{}
\newcommand{\cX}{{\mathcal X}}
\newcommand{\cC}{{\mathcal C}}
\newcommand{\cY}{{\mathcal Y}}
\newcommand{\cZ}{{\mathcal Z}}

\newcommand{\cK}{{\mathcal K}}
\ifCLASSINFOpdf
\else
\fi

\begin{document}
\title{Wiretap Channel With Causal State Information and Secure Rate-Limited Feedback\thanks{Parts of this work appeared at the 51st Annual Allerton Conference on Communication, Control, and Computing, 2013.}}
%\markboth{}{}
\author{\IEEEauthorblockN{Alejandro Cohen and Asaf Cohen}\\
\IEEEauthorblockA{Department of Communication Systems Engineering\\
Ben-Gurion University of the Negev,\\ Beer-Sheva, 84105, Israel\\
Email: {\{alejandr,coasaf\}}@bgu.ac.il}}
\maketitle
\begin{abstract}
In this paper, we consider the secrecy capacity of a wiretap channel in the presence of causal state information and secure rate-limited feedback. In this scenario, the causal state information from the channel is available to both the legitimate transmitter and the legitimate receiver. In addition, the legitimate receiver can send secure feedback to the transmitter at a limited rate $R_f$.

We derive upper and lower bounds on the secrecy capacity and show that, when the channel to the eavesdropper is \emph{degraded}, the bounds are tight and the secrecy capacity is completely characterised.

The capacity achieving scheme is based on Wyner, Csisz\'{a}r and K\"{o}rner wiretap coding and two steps of shared-key generation: one from the state information and one via the noiseless feedback. The upper bound is more involved and requires a non-trivial recursive lemma extending previous results in the literature to include both state and feedback.
We conclude the paper by showing that a few interesting known results can be seen as special cases of the above, especially the case where the state information is available only at the decoder, and the suggested scheme achieves the secrecy capacity without a source of randomness at the decoder.
\end{abstract}

\section{Introduction}\label{intro}
The increasing demand for network connectivity and high data rates dictate efficient utilization of resources, such as the sharing of a common medium for communication. However, in many practical applications, it is important to assure privacy is not compromised. In some systems, cryptographic schemes can be used to protect data from eavesdropping. Yet, these schemes usually involve a computational burden. \emph{Information theoretic security}, on the other hand, offers privacy at the price of transmission rate.

A canonical model in the context of channel coding was given by Wyner in \cite{C2}. Therein, the wiretap channel, described in Figure 1, was introduced. In a wiretap channel, Bob receives the transmitted message of Alice via a channel C1, called the main channel. The eavesdropper Eve, however, observes the information transmitted by Alice through the wiretap channel, C2.
The legitimate parties wish to communicate through C1 while concealing the information from Eve. Specifically, Alice wishes to encode its message $M$ and transmit a codeword $X^n$ on the channel C1. Bob receives $Y^n$, while Eve receives $Z^n$. The legitimate pair's objectives are security, that is, $Z^n$ should provide no information about $M$ (more precisely, $\frac{1}{n}I(M;Z)\to 0$ as $n\to \infty$), and reliability, that is, $M$ should be decoded from $Y^n$ with a negligibly small probability of error. The maximal rate $\frac{1}{n}\log M$ at which both objectives can be fulfilled is called the \emph{secrecy capacity}.

For a \emph{physically degraded} wiretap channel, i.e., when $p(y,z|x) = p(y|x)p(z|y)$, Wyner showed that the secrecy capacity is
\begin{equation}\label{capacity Wyner}
C_{s}=\max_{p(x)}\{I(X;Y)-I(X;Z)\}.
\end{equation}
\begin{figure}
\centering
\includegraphics[scale=1]{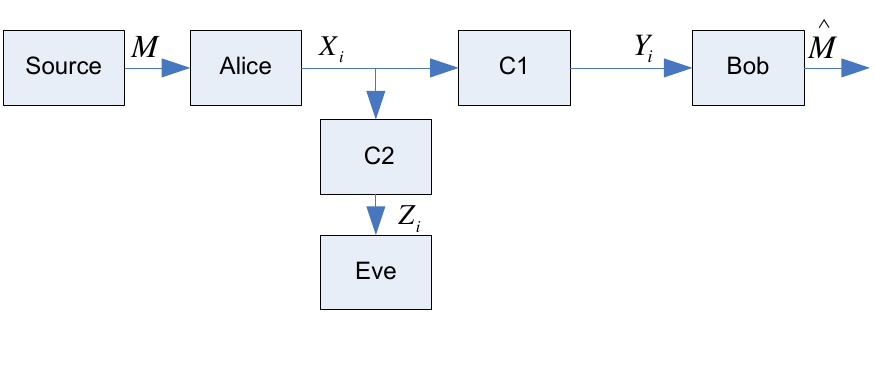}
\caption{Discrete Memoryless Wiretap Channel (DMWTC).}
\label{wiretap channel}
\end{figure}
This result was later extended by Csisz\'{a}r and K\"{o}rner in \cite{C3}, which considered \emph{broadcast channels with confidential messages}. A special case of the results therein is the secrecy capacity of the \emph{general} wiretap channel $p(y,z|x)$, namely,
\begin{equation*}
C_{s}=\max_{p(u,x)}\{I(U;Y)-I(U;Z)\},
\end{equation*}
with an auxiliary $U$ whose cardinality is bounded by that of $X$.
Of course, this reduces to $(1)$ when the wiretap channel is degraded. In fact, it suffices that the main channel is \emph{more capable}, that is $I(X;Y)\geq I(X;Z)$ for any $p(x)$, for the Wyner result in $(1)$ to hold. An important concept in the achieveability of the above results is the \emph{added randomness}, used to confuse the eavesdropper regarding the actual message sent. Such randomness will play a key role in this paper as well.

\begin{figure}
\centering
\includegraphics[scale=0.9]{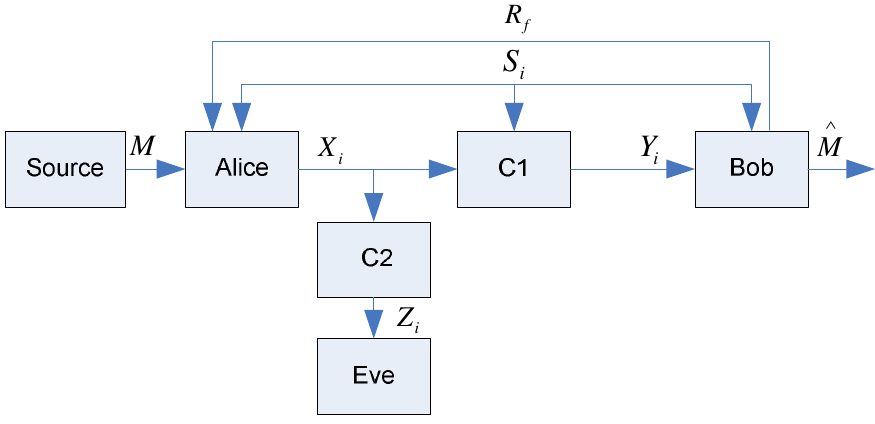}
\caption{The setting considered in this paper. On top of the Wyner wiretap model, a feedback link is available from the legitimate receiver to the transmitter. Moreover, the state information at the main channel is available causally to the legitimate parties. Both the main channel and the wiretap channel are DMCs.}
\label{figure:2}
\end{figure}

The current literature includes several generalizations of the canonical models given in \cite{C2} and \cite{C3}. We include here only the most relevant. A thorough discussion of related works is given in Section III.
In \cite{ahlswede2006transmission}, Ahlswede and Cai considered a discrete memoryless wiretap channel with secure \emph{output feedback}. A general feedback link was considered by Ardestanizadeh \emph{et al.} in \cite{C8}. Therein, since the feedback was not limited to merely pass the output symbols $Y^n$, the authors showed that it is beneficial to use the feedback to send \emph{fresh randomness}, to be used as a shared key between Alice and Bob.

In \cite{C9}, Liu and Chen considered a wiretap model where the main channel is a \emph{state-dependent} DMC. While the eavesdropper remains ignorant of the state, the authors considered the cases where the transmitter and legitimate receiver may or may not have non-causal knowledge of the state. The closely related problem of \emph{secret key agreement} in the wiretap channel with non-causal state information was considered in \cite{khisti2011secret}. The work in \cite{C9} was later extended by Chia and El Gamal in \cite{C11} to the case where \emph{causal} state information is available (i.e., only past and current state values are given). At the heart of Chia and El Gamal's methods stands a key generation scheme. Again, this key, shared by Alice and Bob, is used to increase the secrecy capacity. The works in \cite{C9,khisti2011secret,C11}, however, do not include a feedback link.
%%%%%%%%%%%%%%%%%%%%%%%%%%%%%%%%%
\subsection*{Main Contribution}
In this work, we consider the system depicted in Figure 2. In this setting, both channel state information (CSI) denoted as $S_i$ and a rate limited feedback denoted as $R_{f}$ are available. We derive upper and lower bounds on the secrecy capacity, and show that when the eavesdropper's channel is degraded, namely, $p(y,z|x,s) = p(y|x,s)p(z|y)$, the bounds are tight and describe the secrecy capacity exactly. In the lower bound, we show that a combined scheme of both types of key generation is required to achieve the results: on top of the Wyner scheme, one has to create a shared key from the state information and send an additional key through the feedback. The main contribution is in the upper bound, which is more involved, and requires showing that indeed such a use of the feedback link is optimal. We prove the upper bound via a non-trivial recursive lemma, which enables us to include \emph{both state and feedback}. The resulting region reduces to previously known results in the literature when the feedback or state information are not available, and thus extends them. When the state information is available only at the legitimate decoder, the state information can be viewed as part of the channel output, hence the framework of \cite{C8} applies. However, a comparison of achievable schemes suggests that, unlike what \cite{C8} suggests, one should not use the feedback to send fresh randomness, but might as well \emph{losslessly describe the state to the transmitter}, avoiding the need for randomness at the decoder.

Applications of the above results include, but are not limited to, cases where the eavesdropper channel is \emph{not weaker than the main channel}, yet, one can achieve secure communication via channel state and feedback, extending the range of scenarios where information theoretic security can be used. Moreover, a deep understanding of the capacity of the wiretap channel under diverse conditions such as state information and feedback will facilitate the application of such physical layer security concepts to modern, real word networks, where state information and feedback are available (through estimation or two way communication), but to date, are not used at their full potential.

The structure of this work is as follows. In Section II, the problem is formally described. In Section III, we summarize the related work. Section IV includes our main results, with the lower bound proved in Section V and the upper bound in Section VI. Section VII includes a description of a few important special cases. Section VIII concludes the paper.

\section{Problem Formulation}\label{formulation}
We consider the Discrete Memoryless Wiretap Channel (DMWTC) with secure rate limited feedback and causal state information given in Figure 2. Both Alice, the encoder, and Bob, the decoder, have access to the state information $S_i$. In general, the state information can be available non-causally or causally. In this paper, we focus on the causal case. Alice desire is keeping Eve ignorant of the confidential message, denoted as $M\in \{1,\ldots,2^{nR}\}$, sent to Bob.

Throughout the paper, capital letters denote random variables, lower case letters denote their realizations, and calligraphic letters denote the alphabet. Thus, the sent message is $(X_1,\ldots,X_n)=X^n$, $X \in \cX$, the output at the legitimate receiver is $Y^n$, $Y \in \cY$, and the output at the eavesdropper is $Z^n$, $Z\in \cZ$. The main channel is affected by a \emph{memoryless} state sequence $S^n$, $S\in \mathcal{S}$, known causally to both the encoder and the legitimate decoder. We assume a memoryless channel, that is,
\begin{eqnarray*}
p(y_{i},z_{i}|x^{i},y^{i-1},z^{i-1},s_{i})&=&p(y_{i},z_{i}|x_{i},s_{i}).
\end{eqnarray*}
In the case where just the main channel is affected by the state information, the cross-over probabilities at the wiretap channel is $p(z_{i}|x^{i},z^{i-1})=p(z_{i}|x_{i})$.

We assume a rate-limited feedback at rate $R_{f}$ is available from the decoder to the encoder.
That is, symbols $K^{f}_{i}$, $i \in \{1,\ldots,n\}$ are sent over a feedback link, secretly from the eavesdropper. The feedback alphabets are denoted as $\{\cK_{1}^{f},\ldots,\cK_{n}^{f}\}$. Thus, their cardinalities must satisfy
\begin{equation} \label{model 0}
\frac{1}{n}\sum_{i=1}^{n}\log(|\cK_{i}^{f}|)\leq R_{f}.
\end{equation}
The symbol $K^{f}_{i}$ at instant $i$ may depend on prior outputs up to instant $i-1$ of the channel, $Y^{i-1}=(Y_{1},\ldots Y_{i-1})$ and the prior symbols $(K^{f}_{1},\ldots K^{f}_{i-1})=K_{f}^{i-1}$. Note that, with a slight misuse of notation, a single instant of the feedback value at time $i$ is denoted $K^f_i$ while the \emph{vector} $(K_1^f,\dots,K_i^f)$ is denoted $K^i_f$. This will remain throughout the paper. We allow a random feedback, that is, its actual values can depend on some conditional probability distributions
\[
p(k^{f}_{i}|y^{i-1},k_{f}^{i-1}).
\]
Consequently, a code with parameters $(2^{nR},2^{nR_f},n)$ for the wiretap channel in the presence of causal state information and rate-limited feedback is defined by a message set $\{1,\ldots,2^{nR}\}$; The conditional probability distributions of the stochastic coding for the legitimate encoder
\[
p(x_{i}|m,x^{i-1},s^i,k^{i}_{f}),
\]
where $m$ denotes the message to be sent; The feedback at rate $R_f$ defined above and, finally, a decoding map
\[
\hat{m} : \cY^n \times \mathcal{S}^n \times \cK^{n}_{f} \mapsto \{1,\ldots,2^{nR}\}.
\]
Hence, the decoded message is $\hat{M} = \hat{m}(Y^n,S^n,K^{n}_{f})$.
The message $M$ at the legitimate encoder is distributed uniformly on $\{1,\ldots,2^{nR}\}$, thus $H(M)= n
R$.

The normalized equivocation at the eavesdropper is the ratio $H(M|Z^{n})/H(M)$. Denote the error probability $p(\hat{M} \ne M)$ as $P_e^{(n)}$. We say that the rate/normalized equivocation tuple $(R ,R_f ,d)$ is achievable if for any $\varepsilon >0$ there exists an $(2^{nR},2^{nR_f},n)$ code such that
\begin{equation}\label{conds}
\frac{H(M|Z^{n})}{H(M)}\geq d-\varepsilon
\quad \text{and} \quad
P_{e}^{(n)}\leq \varepsilon.
\end{equation}
Furthermore, we say the secrecy capacity is $C_{s}$ if $C_s$ is the supremum of $R$ in the tuples $(R,R_f,1)$ satisfying the inequalities in $(3)$. Namely, tuples where asymptotically, the eavesdropper is ignorant of the message sent. In this paper, we focus \emph{only on the above case where $d=1$}, that is, we are interested in the secrecy capacity $C_{s}$.

\section{Related Work}\label{related work}
The first information theoretic study on the problem of securely transmitting a message over a public channel was done in \cite{C1}. Therein, Shannon considered the problem of transmitting a message $M$ from the legitimate sender to the legitimate receiver via an open channel. Perfect secrecy was defined by $I(M;X) = 0$, where $X$ is the transmitted word. The result was that the legitimate parties must share a key of the same length as the message itself in order to achieve such a strong secrecy requirement. However, as was shown later, slightly relaxing the perfect secrecy constraint is beneficial.

% general wiretap
The wiretap channel in Figurw 1 was presented in \cite{C2}, under only an \emph{asymptotic} independence constraint, requiring $\frac{1}{n}I(M;Z^n)$ to vanish. For this channel, one would expect that when the capacity of the main channel is larger than that of the wiretap channel, the secrecy capacity will be positive. Indeed, in \cite{C2} Wyner proved that when the wiretap channel is a degraded version of the main channel, the secrecy capacity is positive. \cite{C3} extended this and concluded that a positive secrecy capacity is possible whenever the main channel is \emph{less noisy} than the wiretap, that is, less noisy than $Z$, namely, $I(U;Y) \geq I(U;Z)$ for any auxiliary $U$ such that $U\leftrightarrow X\leftrightarrow (Y,Z)$. Moreover, as was later discussed in ~\cite{C9,C8,C11}, even if the wiretap channel is not degraded, or less noisy, the secrecy capacity can be positive using \emph{state information}. We elaborate on this now.
%%%%%%%%%%%%%%%%%%%%%%%%%%%%%%%%%%%%%
\subsection{Channel State Information}\label{related state}
In \cite{C9}, a DMWTC with CSI at both ends was discussed. Specifically, the CSI $S^{N}_{1}$ was given non-causally at the encoder while $S^{N}_{2}$ was given non-causally at the decoder. The main channel was specified through $p(y|x,s_1,s_2)$.
The following information rates were defined.
\begin{eqnarray*}
R_{U1}&=&I(U;Y,S_{2})-\max\{I(U;Z),I(U;S_{1})\}\label{W. Liu and B. Chen 1},\\
R_{U2}&=&I(U;Y,S_{2})-I(U;S_{1}).
\end{eqnarray*}
$U$ was an auxiliary random variable and, in addition, a Markov condition $U \leftrightarrow (X, S_{1}, S_{2}) \leftrightarrow (Y,Z)$ must hold. It was shown that the set
\begin{equation}\label{W. Liu and B. Chen 2}
R_{s}=\bigcup_{U \leftrightarrow (X, S_{1}, S_{2}) \leftrightarrow (Y,Z)} (R,d),
\end{equation}
is achievable, under the constraints $Rd\leq R_{U1}$ ($0\leq d\leq 1$) and $0\leq R\leq R_{U2}$. Of special interest to us is the case where $d=1$ and we have
\[
R \leq \max_{U \leftrightarrow (X, S_{1}, S_{2}) \leftrightarrow (Y,Z)}\min\{R_{U1},R_{U2}\}.
\]
This is an achievable rate for the wiretap channel with two-sided non-causal state information.
Moreover, when the state information is available only at the receiver, that is $S_{1}=\emptyset$ and $S_{2}=S$, we have
\begin{eqnarray*}
R_{U1}&=&I(U;Y,S)-I(U;Z),\\
R_{U2}&=&I(U;Y,S).
\end{eqnarray*}

In a recent paper \cite{C11}, Chia and El-Gamal considered a similar setting, yet with causal state information. As mentioned, at the heart of the scheme is a key generation step. In particular, random binning of the state sequence known to both the encoder and decoder gives rise to a \emph{shared key}. This key is used to encrypt part of the message. The achievable scheme results in higher rates compared to \cite{C9}. In particular, they established that
\begin{eqnarray}\label{eq:EL_Gamal_1}
C_{S} &\geq&\max\{\max_{p(u|s)p(x|u,s)}\min \{I(U;Y|S)-I(U;Z|S)\nonumber\\
&&+H(S|Z),I(U;Y|S)\},\nonumber\\
&& \max_{p(u)p(x|u,s)} \min\{H(S|Z,U),I(U;Y|S)\}\}.
\end{eqnarray}
When $Y$ is less noisy than $Z$, namely, $I(U;Y|S) \geq I(U;Z|S)$ for any auxiliary $U$ such that $(U,S)\leftrightarrow (X,S)\leftrightarrow (Y,Z)$, this results in
\begin{multline*}
C_{S}  =  \max_{p(x|s)}\min \{I(X;Y|S)-I(X;Z|S)
 +  H(S|Z),\\I(X;Y|S)\}.
  \end{multline*}

% more with state
Specific results for the Gaussian channel were given in \cite{leung1978gaussian} and in \cite{C14} for a Gaussian channel with state.
Secret agreement schemes for channels with state known at the transmitter were given in \cite{khisti2011secret}. Of course, CSI plays a key role in MIMO-Wiretap channels as well. Secrecy capacity (and in particular, outage) was discussed in \cite{gerbracht2012secrecy}, where it was shown that one can take the advantage of having perfect knowledge of the channel state to the legitimate receiver by adding artificial noise to the null space of the main channel. Later results regarding the optimal beamforming vectors can be found in \cite{li2012optimality}. The advantage of perfect knowledge of the channel state to the legitimate receiver was also used in \cite{li2011ergodic} under a Rician fading setting. Fading channels in the context of OFDM were considered in \cite{renna2012physical}. Finally, the current literature also includes work on \emph{active eavesdropping} in the presence of channel state \cite{boche2012comparison}.
%%%%%%%%%%%%%%%%%%%%%%%%%%%%%%%%%%%%%%%%%
\subsection{Feedback}
In this work, besides state information, we are interested in the availability of noiseless feedback as well. Interactions in secure duplex communication can benefit from the insights of such models. Indeed, we will establish that in the case of wiretap channels with CSI, feedback can increase the secrecy capacity significantly.

In \cite{ahlswede2006transmission}, the authors characterized the capacity of a wiretap channel with \emph{noiseless feedback of the channel output}, and showed that in this case,
\begin{equation}\label{output feedback}
C_s = \max_{p(x)} \min \{I(X;Y|Z) + H(Y|X,Z),I(X;Y) \}.
\end{equation}
This is exactly the capacity when one extracts a key from the feedback (as feedback does not increase capacity otherwise).

In ~\cite{C8}, the authors investigated the secrecy capacity $C_{s}(R_{f})$ of a wiretap channel where the legitimate parties have a secure link (feedback) from the decoder to the encoder \emph{at rate $R_{f}$}. Note that the feedback was not limited to carry output symbols. The upper bound in \cite{C8} was
\begin{equation}\label{eq. only feedback}
C_{s}(R_{f})\leq \max_{p(x)}\min\{I(X;Y|Z)+R_{f},I(X;Y)\}.
\end{equation}
Again, when the wiretap channel is physically degraded, the upper bound is tight, establishing $(7)$ as the secrecy capacity.

% practical schemes and networks
\subsection{Secret Key Agreement, Networks and Practical Schemes}\label{scg and schemes}
A closely related set of problems fall under \emph{secret key agreement} schemes \cite{maurer1993secret,ahlswede1993common,ahlswede1998common}. In these problems, one is interested in the \emph{number of secret bits} which can be distilled in a system. In fact, it is not hard to see that any wiretap coding scheme is also a secret key agreement scheme, hence the secret key capacity is an upper bound on the secrecy rate. However, it is tight only in specific cases (e.g., \cite[Example 4.7]{bloch2011physical}). The difference stems from the fact that the underlying metric is different (number of secret bits distilled versus number of bits transmitted securely). Moreover, many secret key agreement schemes allow public information exchanges between Alice and Bob (on top of the main channel or the correlated sources available to them). To the best of our knowledge, there are no concrete results from the literature on secret key capacity which, for the problem at hand, give tighter bounds than the ones we give in the paper.

The theory and practice of wiretap channel coding has been found useful in a variety of networking scenarios as well. E.g., in \cite{bloch2008network}, the authors suggested a client-server setting where the binary erasure wiretap channel approach was found constructive. A multitude of network coding scenarios exist in \cite{cai2011secure,el2012secure}.

Finally, we mention that the current literature includes practical coding schemes as well, rendering them as desirable solutions for physical layer security. LDPC for the Gaussian wiretap channel was suggested in \cite{klinc2011ldpc}, as well as in \cite{wong2011secret} to suggest secret sharing schemes. LDPCs were also used to allow strong secrecy over a binary erasure channel in \cite{subramanian2011strong}.

\section{Main Results}\label{main results}
We list the main results of the paper, and conclude with an illustrative example. The proofs are deferred to Sections \ref{direct} and \ref{converse}.
%%%%%%%%%%%%%%%%%%%%%%%%%%%%%%%%%
\subsection{Lower Bound}
The achievability part is given by the following theorem. 
\begin{theorem}\label{direct theorem}
Assume a DMWTC, with causal CSI given at the encoder and the legitimate decoder and in the presence of rate limited feedback. Then, the secrecy capacity is lower bounded by
\begin{multline*}%\label{eq:main results_1}
C_{S}  \geq \max\{\max_{p(u|s)p(x|u,s)}\min\{I(U;Y|S)-I(U;Z|S)\\
 + H(S|Z)+R_{f},I(U;Y|S)\},\\
 \max_{p(u)p(x|u,s)}\min\{H(S|Z,U)+R_{f},I(U;Y|S)\}\},
\end{multline*}
where $U$ is an auxiliary random variable and the distributions $p(u|s)$ and $p(x|u,s)$ can be optimized.
\end{theorem}
Theorem 1  proves that the secrecy capacity is dictated by the following three rates:
\begin{eqnarray*}
\hat{R}_{1} & = & I(U;Y|S)-I(U;Z|S)+H(S|Z)+R_{f},\\
\hat{R}_{2} & = & I(U;Y,S)-I(U;S) = I(U;Y|S),\\
\hat{R}_{3} & = & H(S|Z,U)+R_{f}.
\end{eqnarray*}
Note that $\hat{R}_{3}$ is the maximal rate at which a key can be extracted and used to secure the communication. Clearly, up to this rate, there is no need in Wyner-like wiretap coding, and the message can be protected solely by the key. On the other hand, $\hat{R}_{2}$ is the capacity of the main channel. This bound certainly cannot be exceeded. Between these two bounds, Wyner-like coding is beneficial, and the resulting secrecy rate will be $\hat{R}_{1}$, which is the most interesting constraint. 

The achievability, while based on extracting a key from the side information sequence similar to \cite{C11}, involves both Wyner-like wiretap coding and \emph{two levels of protection} using secret keys: one extracted from the CSI and one transmitted through the feedback. Hence, a message is divided into \emph{three sub-messages}, which should not leak to the eavesdropper. However, \emph{as feedback introduces memory, a key challenge is in proving a corresponding upper bound}. 
%%%%%%%%%%%%%%%%%%%%%%
\subsection{Upper Bound}
The upper bound is the following.
\begin{theorem}\label{converse theorem}
Assume a DMWTC, with causal CSI given at the encoder and the legitimate decoder and in the presence of rate limited feedback. Then, the secrecy capacity is upper bounded by
\begin{multline*} %\label{eq:R case 5 23}
C_{S}  \leq  \max_{p(x|s)}\min\{I(X;Y|Z,S)\\
 +H(S|Z)+R_{f},I(X;Y|S)\}.
\end{multline*}
\end{theorem}
The availability of a rate-limited feedback complicates the converse as it renders many of the tools used for memoryless channels impractical. Specifically, in the presence of feedback, $p(y^n,z^n|x^n,s^n)$ no longer has a simple product form. Thus, the proof of the converse involves a non-trivial generalization of a recursive lemma from \cite{C8}, which incorporates both state information and feedback.
%%%%%%%%%%%%%%%%%%%
\subsection{Complete Characterization of the Secrecy Capacity}\label{complete characterization}
It is interesting to investigate the cases where the lower and upper bounds match.
As for the lower bound, when $\hat{R}_{1} \geq \hat{R}_{3}$ \emph{and the main channel is less noisy than the eavesdropper channel}, with the data processing inequality $I(U;Y|S) \leq I(X;Y|S)$ the bound sums up to
\[
C_{S}  \geq  \max_{p(x|s)}\left\{I(X;Y|S)-I(X;Z|S)+ H(S|Z)+R_{f}\right\}.
\]
Note that the less noisy assumption is weaker than the degraded assumption, hence the result holds in this case as well. Now, when the eavesdropper channel is degraded such that $p(y,z|x,s) = p(y|x,s)p(z|y)$ and $(X,S)\leftrightarrow Y \leftrightarrow Z$ from a Markov chain, the upper bound on the secrecy capacity reduces to
\[
C_{S}  \leq  \max_{p(x|s)}\left\{I(X;Y|S)-I(X;Z|S)+ H(S|Z)+R_{f}\right\}
\]
and the bounds match.
\begin{corollary}
Assume a DMWTC, with causal CSI given at the encoder and the legitimate decoder and in the presence of rate limited feedback. If the eavesdropper channel is degraded, then the secrecy capacity is
\begin{multline*}%\label{eq:Secrecy Capacity Results 1}
C_{S}  =  \max_{p(x|s)}\min\{I(X;Y|S)-I(X;Z|S)\\
 + H(S|Z)+R_{f},I(X;Y|S)\}.
\end{multline*}
\end{corollary}
%%%%%%%%%%%%%%%%%%%%%%%%%%%%%

\subsection{Example}
We compare several scenarios for a degraded binary symmetric wiretap channel, where the scenarios differ by the availability of channel state and noiseless feedback.

A degraded DMWTC with noiseless feedback and without CSI is given in Figure 3  A. The main channel between Alice and Bob is a binary symmetric channel with a transition probability $p_y$. That is,
\begin{eqnarray}
p_{y|x}(0|0)=1-p_{y},\quad p_{y|x}(0|1)=p_{y}\nonumber\\
p_{y|x}(1|1)=1-p_{y},\quad p_{y|x}(1|0)=p_{y}\nonumber.
\end{eqnarray}
We denote this channel model by $BSC(p_y)$.
The wiretap channel is realized by cascading the main binary symmetric channel, $BSC(p_{y})$, and the eavesdropper's binary symmetric channel, $BSC(p_{z})$.

A degraded wiretap channel with noiseless feedback and CSI is shown in Figure 3  B. Now, the main BSC is state-dependent, that is, it is $BSC(p_{s_{i}})$ if the state at time $i$ is $s_i$. The wiretap channel is a cascade of this channel and $BSC(p_{z})$.
The maximum of $C_{s}^{NS}(R_{f})$ (over the input distribution), i.e., the capacity of the degraded DMWTC without CSI yet with feedback, and the maximum of $C_{s}^{S}(R_{f})$, i.e., the capacity of the degraded DMWTC with CSI and feedback, are both achieved using a symmetric input probability distribution, namely, $P(X=0)=P(X=1)=0.5$. Thus, $X$, $Y$ and $Z$ are uniformly distributed over $\{0,1\}$. The binary entropy function and the binary convolution are denoted by $h(p)=-p\log p-(1-p)\log(1-p)$ and by $p_{y}\ast p_{z} = p_{y}(1-p_{z})+(1-p_{y})p_{z}$, respectively. With these distributions, without state information (case A) the mutual information is
\[
I(X;Y)=1-h(p_{y}),
\]
hence the relevant capacity is
\[
I(X;Y)-I(X;Z)=h(p_{y}\ast p_{z})-h(p_{y}).
\]
Now, assume the main channel has state (Case B). We assume only two states are possible, and each state corresponds to a different cross over probability in the main channel. That is, assume $P(S=s_0)=1-q$ and $P(S=s_1)=q$. When $S=s_0$, $p_{y|x,s}(y|x,0)=1-p_{s_{0}}$ if $x=y$ and $p_{s_{0}}$ otherwise.
When $S=s_1$, $p_{y|x,s}(y|x,1)=1-p_{s_{1}}$ if $x=y$ and $p_{s_{1}}$ otherwise. Hence, for the case with state information, we have
\[
I(X;Y|S) =  1-(1-q)h(p_{s_{0}})-qh(p_{s_{1}}).
\]
Furthermore, with the symmetric input distribution as in this example, it is not hard to verify that $I(Z;S)=0$. Thus,
\[
H(S|Z)=H(S)=h(q).
\]
As for the wiretap channel, we have
\[
I(X;Z|S)=1-(1-q)h(p_{s_{0}}\ast p_{z})-qh(p_{s_{1}}\ast p_{z}).
\]
Hence,
\begin{align*}
I(X;Y|S)&-I(X;Z|S)+H(S|Z)\nonumber\\
=&1-(1-q)h(p_{s_{0}})-qh(p_{s_{1}})\nonumber\\
&-\{1-qh(p_{s_{1}}\ast p_{z})-(1-q)h(p_{s_{0}}\ast p_{z})\}
\\ &\quad +h(q)\nonumber\\
=&(1-q)h(p_{s_{0}}\ast p_{z})+qh(p_{s_{1}}\ast p_{z})\nonumber\\
& -(1-q)h(p_{s_{0}})-qh(p_{s_{1}})+h(q).
\end{align*}
\begin{figure}
  \centering
  \includegraphics[scale=1]{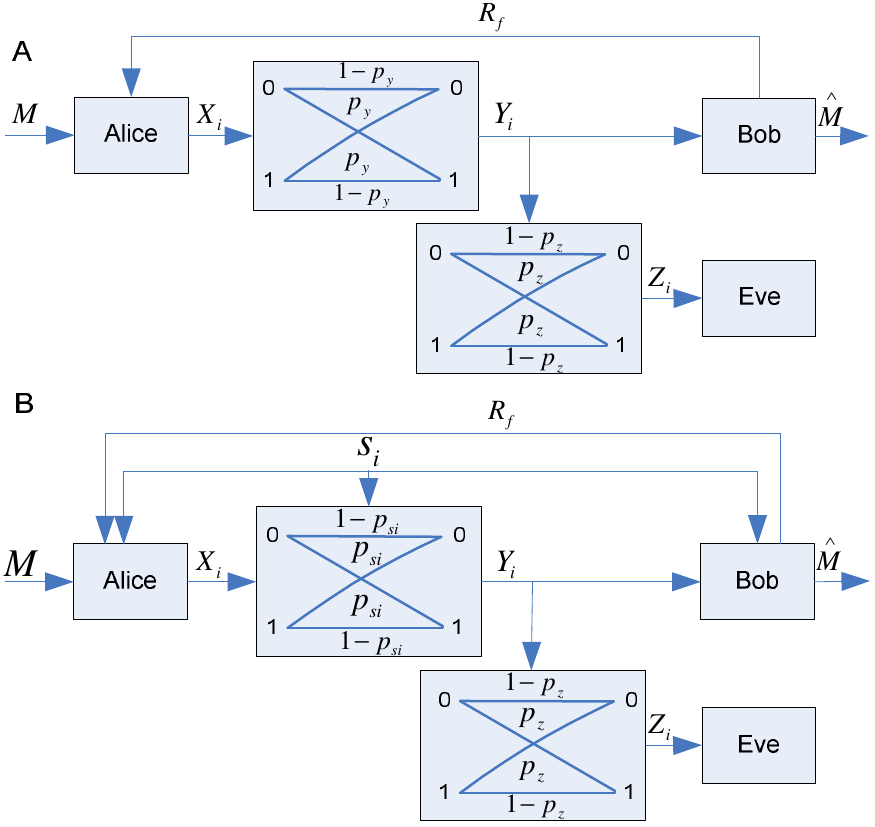}
  \caption{(A) Degraded wiretap channel without causal CSI, (B) Degraded wiretap channel with causal CSI.}
  \label{figure:4}
\end{figure}
As a result, the capacities to compare are
\[
C_{s}^{NS}(R_{f})=\min \{1-h(p_{y}),h(p_{y}\ast p_{z})-h(p_{y})+R_{f}\}
\]
and
\begin{multline*}
C_{s}^{S}(R_{f})=\min \{1-(1-q)h(p_{s_{0}})-qh(p_{s_{1}}),\\
(1-q)h(p_{s_{0}}\ast p_{z})+qh(p_{s_{1}}\ast p_{z})\nonumber\\
-(1-q)h(p_{s_{0}})-qh(p_{s_{1}})+h(q)+R_{f}\}.
\end{multline*}

\begin{figure}
  \centering
  \includegraphics[scale=0.45]{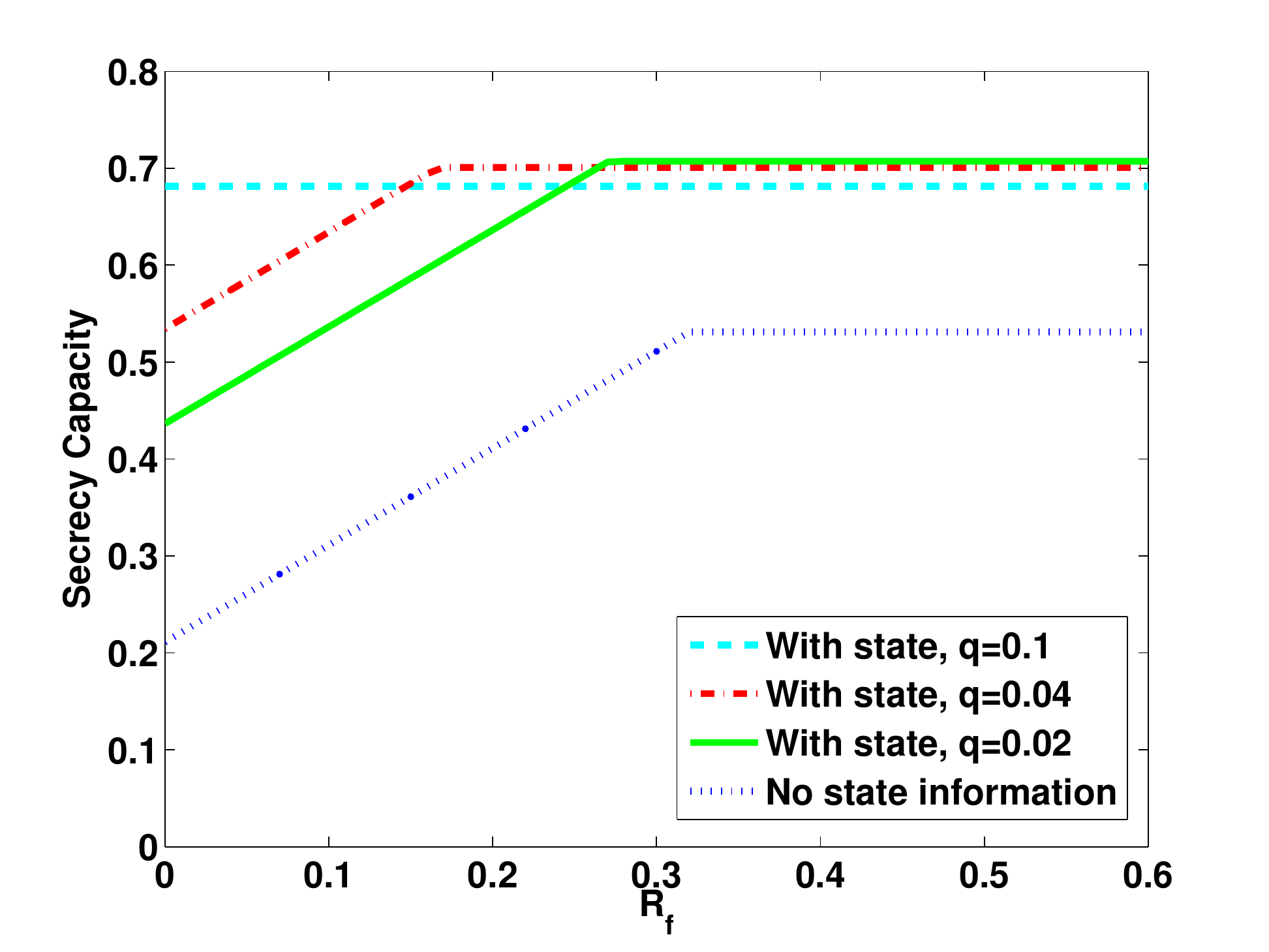}
  \caption{The secrecy capacities $C_{s}^{NS}(R_{f})$ and $C_{s}^{S}(R_{f})$ versus $R_f$, with and without a state sequence. $(q,1-q)$ is the (memoryless) probability distribution for the state sequence. Note that for $q=0.1$, in this case, the amount of randomness in the key suffices to saturate the secrecy capacity regardless of the value of $R_f$. For lower $q$, however, there is not enough randomness in the key generated from the state, and the capacity saturates only for non-zero values of $R_f$.}
  \label{fig:rf}
\end{figure}

\begin{figure}
  \centering
  \includegraphics[scale=0.45]{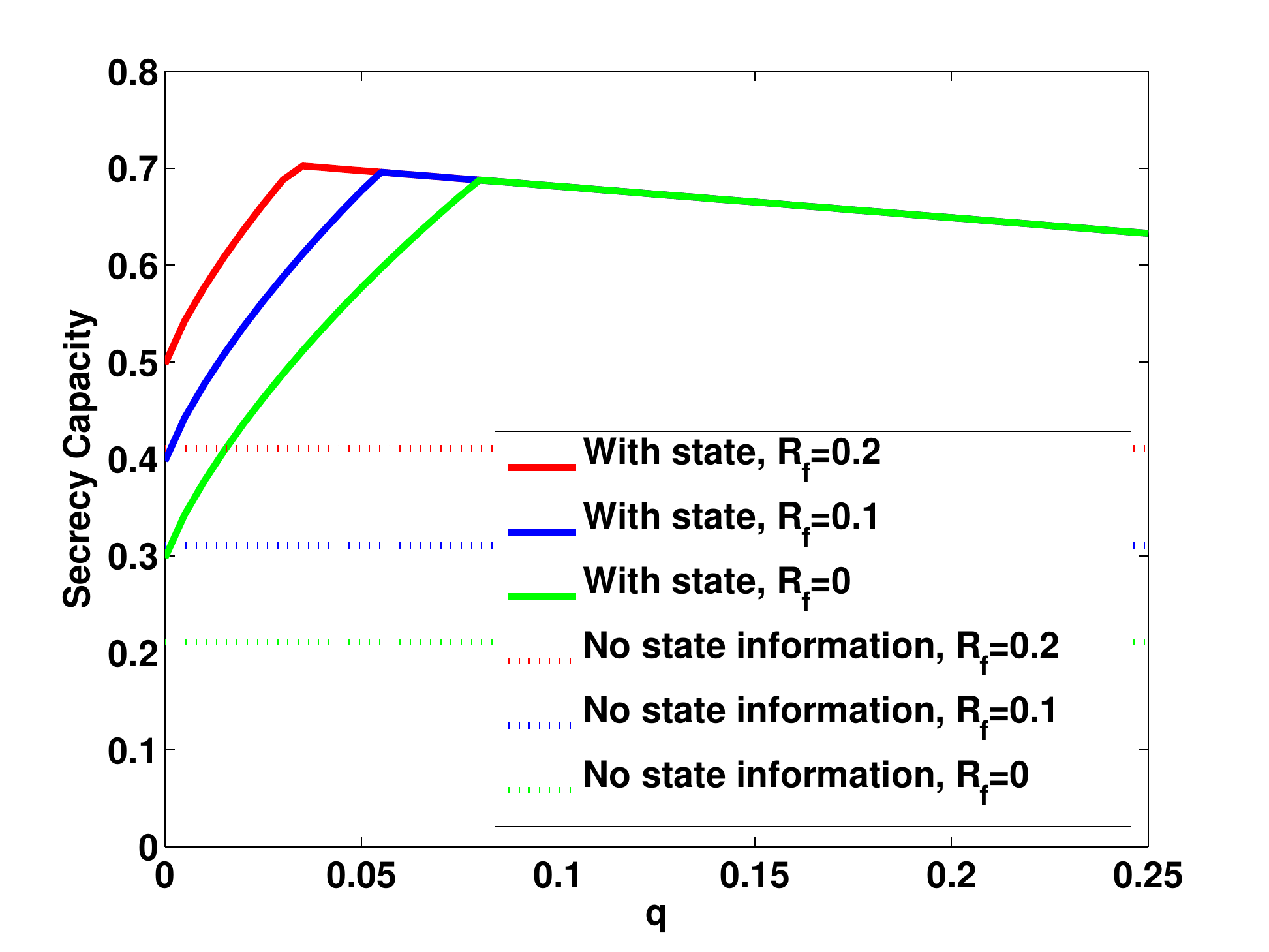}
  \caption{The secrecy capacities, this time as a function of the state probabilities, that is, $C_{s}^{NS}(q)$ and $C_{s}^{S}(q)$. For low values of $q$, the key factor in the secrecy capacity is the amount of randomness in the state sequence. Hence, in this range, the higher $q$ is, the higher the secrecy capacity. However, when the secrecy capacity saturates, due to the specific choices of probabilities $p_{s_0}$ and $p_{s_1}$, the higher $q$ is, lower the main channel capacity, hence the secrecy capacity decreases.}
  \label{fig:q}
\end{figure}

Numerical results for the capacities above are given in Figures \ref{fig:rf} and \ref{fig:q}. The cross-over probabilities are fixed on $p_{z}=p_{y}=0.1$, $p_{s_{0}}=0.05$ and $p_{s_{1}}=0.15$. \ref{fig:rf} gives the capacities $C_{s}^{NS}(R_{f})$ and $C_{s}^{S}(R_{f})$ versus $R_f$. The same saturation phenomenon observed with no state is visible with state as well (with linear increase until the saturation point), however, it is clear that the probabilities for each state affect the capacity. This effect is, in fact, twofold. First, the value of the state affects the capacity of the direct channel. Yet, besides this effect, the fact that \emph{there is a state sequence} affects the secrecy capacity. The more entropy the state sequence has, the higher the key rate, hence the faster the capacity saturates at its maximal value (which is the main channel capacity). \ref{fig:q} gives another perspective, plotting the secrecy capacities as a function of $q$. It is clear that once the capacities saturate, the value is independent of $R_f$. It decreases with $q$, though the higher $q$ is the lower is the capacity of the direct channel. This is only due to the specific choices of $p_{s_0}$ and $p_{s_1}$.

\section{Lower Bound}\label{direct}
%This section, we prove our direct result.
%%%%%%%%%%%%%%%%%%%%%%%%%%%%%%%%%%%%%%%%%%%%%%%%
\subsection{Achievability of $\hat{R}_{1}$}\label{sec. r1}
Consider $\hat{R}_{1}$, with the maximization on the input distribution:
\begin{multline*}
\hat{R}_{1}=\max_{p(u|s)p(x|u,s)}\{I(U;Y|S)-I(U;Z|S)\\+H(S|Z)+R_{f}\}.
\end{multline*}
Similar to \cite{C11}, we perform the maximization in $\hat{R}_{1}$ through distributions $p(u'),p(x|u,s)$ and functions of the form $u(u',s)$, using the functional representation lemma \cite{C15}. This way, the achievability can be proved for an equivalent characterization of $\hat{R}_{1}$,
\begin{multline*}
\max_{p(u'),u(u',s),p(x|u,s)}\{I(U';Y,S)-I(U';Z,S)\\
 +H(S|Z)+R_{f}\}.
\end{multline*}

We split the proof to two cases, the first is when $I(U';Y,S)\geq I(U';Z,S)$, where for this case $(U',S) \leftrightarrow (X,S)\leftrightarrow(Y,Z)$ from a Markov chain, and the second case is when $I(U';Y,S)\leq I(U';Z,S)$.
%%%%%%%%%%%%%%%%%%%%%%%%%%%%%%%%%%%%%%%%%%%
\subsubsection{First Case - $I(U';Y,S)\geq I(U';Z,S)$}
\paragraph{Encoding of Legitimate Sender (Alice)}
The encoding scheme in this case requires the transmission of $B - 1$ protected blocks during the transmission of $B$ blocks, each of length $n$. The message in the first block is not fully protected.

Given a distribution $P_{U'}$ and a function $u(u',s)$, we set the following three rates:
\begin{eqnarray*}
R_0 &=& I(U';Y,S)-I(U';Z,S) - 2\epsilon,
\\
R_1 &=& H(S|Z) - \epsilon,
\\
R_2 &=& R_f.
\end{eqnarray*}
The coding scheme corresponds to a transmission rate
\[
R = R_0 + R_1 + R_2 = \hat{R}_{1} - 3 \epsilon,
\]
secretly from the eavesdropper. 
We assume $R \leq I(U'; Y,S)$.
Of course, this is not an actual restriction as $I(U'; Y,S) \leq I(U'; Y|S)$, which is, in turn, an upper bound on the secrecy capacity \ref{direct theorem} claims is achievable. Hence, cases where $R_0 + R_1 + R_2 > I(U'; Y,S)$ are of no interest.

We split the message $M_{j},j\in \{2,\ldots,B\}$, into three independent messages $M^{0}\in \{1,\ldots,2^{nR_{0}}\}$, $M^{1}\in \{1,\ldots,2^{nR_{1}}\}$ and $M^{2}\in \{1,\ldots,2^{nR_{2}}\}$. The message at rate $R_0$ will be protected by the Wyner wiretap coding scheme, while the messages at rate $R_1+R_2$ will be protected by the keys: a message at rate $R_1$ protected by the key generated from the CSI and a message at rate $R_2$ protected by the key received from the feedback.

The first step is the generation of the message codebook. We randomly generate $2^{n[I(U';Y,S)-\epsilon]}$ i.i.d.\ sequences $u'^{n}(l)$, using the distribution $P(U'^{n}= u'_{i})=\prod_{i=1}^{n}P_{U'}(u'_{i})$.
Then, these sequences are distributed randomly into $2^{nR_{0}}$ equal size bins. The index of each bin is denoted as $j\in \{1, 2,\ldots J=2^{nR_0}\}$. Next, these sequences are distributed randomly into $2^{nR_{1}}$ sub-bins, and we further partition each sub-bin to $2^{nR_{2}}$ equal size sub-bins. Denote the resulting bin indices by $C(m^{0},m^{1},m^{2})$.

Next we create the keys from the CSI. We bin the channel state sequences $s^{n}$ at random into $2^{nR_1}$ bins $\{B(k_{s})\}_{k_s=1}^{2^{nR_1}}$. The key $K^{S}_{j-1}$ used to protect $M^1$ in block $j$ is the bin index of the state sequence $S^{n}(j-1)$ in block $j-1$. 

The third step is the generation of the feedback codebook. Similar to \cite{C8}, it is used solely to give the encoder random bits. Herein, however, such random bits are sent for each block. The key is of rate $R_{2}$, i.e., Bob sends $k^{f}_{j}$ drawn uniformly from $2^{nR_{2}}$ indices to be used in the $j$th block. This key is the one used to encrypt $M^2$ of the given block.

To encode the first block of the message $M_{1}$, given $M_{1}^{0}$, $M_{1}^{1}$ and $M_{1}^{2}$, the encoder selects a random codeword $u'^{n}(L)$ from $C(M_{1}^{0},M_{1}^{1},M_{1}^{2})$. Then it computes $u_{i}=u(u'_{i}(L),s_{i})$ and the symbol transmitted is a random one, according to $X_{i}\sim p(x_{i}|s_{i},u_{i})$ for $i \in \{1\ldots n\}$.
Note that the first block is not protected by the keys.
However, during the transmission of the $j-1$ block, the encoder (Alice) gathers two keys, $k^{s}_{j-1}$ from the state sequence and  $k^{f}_{j-1}$ from the feedback. To encode the $j$-th block of the message $M_{j},j\in \{2,\ldots,B\}$, given $M_{j}^{0}$, $M_{j}^{1}$ and $M_{j}^{2}$, the encoder selects a random codeword $u'^{n}(L)$ from $C(M_{j}^{0},M_{j}^{1}\oplus k^{s}_{j-1},M_{j}^{2}\oplus k^{f}_{j-1})$. $\oplus$ denotes modulo-$[2^{R_{1}}]$ and $[2^{R_{2}}]$ additions. It then computes $u_{i}=u(u'_{i}(L),s_{1i})$ and the symbol transmitted is, again, random, according to $X_{i}\sim p(x_{i}|s_{i},u_{i})$ for $i \in \{(j-1)n+1\ldots jn\}$.

Note that $M^{0}$, similar to \cite{C11}, is protected using the Wyner coding scheme, therefore the eavesdropper cannot comprehend this part from the message when $I(U';Y,S_{1})-I(U';Z,S_{2})>0$. The second part, $M^{1}$, is encrypted with the key $k^{s}_{j-1}$ and the third part, $M^{2}$, is encrypted with the key $k^{f}_{j-1}$.
%%%%
\paragraph{Decoding at legitimate receiver (Bob)}
The decoding involves standard joint typicality arguments. We list here only the most important steps.

In the first block the legitimate receiver searches for a word $u'^{n}(L)$ in the codebook, such that $(u'^{n}(l),y^{n}(j),s^{n}(j))$ is jointly typical, then the legitimate receiver (Bob) declares the index of the bin containing this $u'^{n}(l)$ as the message received.

As the number of originally drawn sequences, $2^{n[I(U';Y,S)-\epsilon]}$, is similar to that in \cite{C11}, the analysis of the error probability is similar, and the jointly typical $u'^{n}(l)$ is identified with high probability. The probability that a different sequence is identified is arbitrarily close to zero. Thus, the message indices $m_j^0,m_j^1$ and $m_j^2$ are decoded correctly with high probability.

As for the decoding at the $j$-th block $j\in \{2,\ldots,B\}$,
the decoder uses a similar procedure to retrieve $m_{j}^{0}$, $m_{j}^{1}\oplus k^{s}_{j-1}$ and $m_{j}^{2}\oplus k^{f}_{j-1}$. It then uses $k^{s}_{j-1}$ to retrieve $m_{j}^{1}$ and $k^{f}_{j-1}$ to retrieve $m_{j}^{2}$. It is easy to verify that a rate $\hat{R}_{1}=\min\{I(U';Y|S)-I(U';Z|S)+H(S|Z)+R(f)-\delta_{n},I(U';Y|S)\}$ can be achieved.
%%%%%
\paragraph{Information Leakage at the Eavesdropper (Eve)}
For $j\in \{1,\ldots,B\}$ we let $Z^{n}_{j}$ denote the eavesdropper's observation in block $j$. The information leaked $L(\tilde{C}_{nB})$, given the codebook and coding procedure $\tilde{C}_{nB}$ is then
\begin{align}
\frac{1}{nB}&L(\tilde{C}_{nB}) = \frac{1}{nB}I(M^{0}_{1}\ldots M^{0}_{B}M^{1}_{1}\ldots M^{1}_{B}\nonumber\\
& \hspace{3cm}M^{2}_{1}\ldots M^{2}_{B};Z^{n}_{1}\ldots Z^{n}_{B}|\tilde{C}_{nB})\nonumber\\
&= \frac{1}{nB}I(M^{0}_{1}\ldots M^{0}_{B}M^{1}_{1}\ldots M^{1}_{B};Z^{n}_{1}\ldots Z^{n}_{B}|\tilde{C}_{nB})\nonumber\\
&  \hspace{1cm}+ \frac{1}{nB}I(M^{2}_{1}\ldots M^{2}_{B};Z^{n}_{1}\ldots Z^{n}_{B}\nonumber\\
& \hspace{3cm}|M^{0}_{1}\ldots M^{0}_{B}M^{1}_{1}\ldots M^{1}_{B},\tilde{C}_{nB})\nonumber\\
\end{align}
where the equality is due to the chain rule for mutual information. We now consider the two summands. 
As for the first, this is exactly the information leakage on the messages protected by the Wyner wiretap scheme and the messages protected by the key \emph{drawn from the state sequence}. Hence, by the results of Chia and El Gamal \cite{C11}, specifically, Proposition 1 therein and the discussion which follows, as long as the key extracted from the \emph{state} is at rate smaller than $H(S|Z)$, Eve's knowledge on the key is negligible, hence the first summand is negligible. 

Consider the second summand. $\{M^2_j\}$ are the portions of the messages \emph{protected by the keys received from the feedback}. The keys sent through the feedback are random, independent of all other variables in our problem. Remember that the encoder selects a random codeword from $C(M_{j}^{0},M_{j}^{1}\oplus k^{s}_{j-1},M_{j}^{2}\oplus k^{f}_{j-1})$. Thus, the actually transmitted codewords (at all blocks), and, of course, the received ones at Eve's side may depend on $M_j^2$ only through $M_{j}^{2}\oplus k^{f}_{j-1}$, that is, they are completely independent of $M_j^2$ unless the key is given. 
Due to the above, the second summand is zero. In other words, if one considers $M^{2}_{1}\ldots M^{2}_{B}$ as a single message at rate $B R_2$ and $Z^{n}_{1}\ldots Z^{n}_{B}$ as an output block of size $Bn$, the second summand is equivalent to the expression $I(M_2;Z^n|\cC,M_1)$ that appears in \cite[equation (45)]{C8}, which is shown to be zero therein. We conclude that $\hat{R}_{1}$ is achievable.
%%%%%%%%%%%%%%%%%%%%%%%%%%%%%%%%%%%%
\subsubsection{Second Case - $I(U';Y,S)\leq I(U';Z,S)$}
Herein, the main channel capacity is too low, and the encoder cannot use the Wyner wiretap scheme to secretly send message to the legitimate receiver. Therefore, only the two keys, the one resulting from the CSI and the one resulting from the feedback can be used to protect the message. We only consider the scenario where $(I(U';Y|S)-I(U';Z|S))+(H(S|Z)+R_{f})>0$. Otherwise the secrecy capacity is zero. 

The same key splitting as in \cite{C11} is used. In short, in the block $j \in\{2\ldots B\}$, we split the message to three parts as before, yet protect the first two with the state key. The state information key $k^{s}_{j-1}$ is split to two independent parts, $k^{s}_{(j-1,0)}$ and $k^{s}_{(j-1,1)}$ at rates which coincides with case 2 of $R_{S-CSI-1}$ in ~\cite{C11}. Thus, compared with the first case of $\hat{R}_{1}$, to send message $M_{j},j\in \{2\ldots B\}$, transmit $X^{n}(k^{s}_{(j-1,0)},M_{j}^{1}\oplus k^{s}_{(j-1,1)},M_{j}^{2}\oplus k^{f}_{j-1})\in C_{n}$. 
The reminder of the proof is very similar to \cite{C11}.
%%%%%%%%%%%%%%%%%%%%%%%%%%%%%%%%%%%%%%%%%%%%%%%%%%%%%%%%%%%%%%%%%%%%%%
%%%%%%%%%%%%%%%%%%%%%%%%%%%%%%%%%%%%%%%%%%%%%%%%%%%%%%%%%%%%%%%%%%%%%%
\subsection{Achievability of $\hat{R}_{3}$}
In this case, to encode the message we use purely the key $K_{s}$ in the block, which coincides with $R_{S-CSI-2}$ in Chia and  El Gamal ~\cite[Theorem 1]{C11}. We split the message $M_{j},j\in \{2,\ldots,B\}$, into two independent messages $M^{1}\in \{1,\ldots,2^{nR_{1}}\}$ and $M^{2}\in \{1,\ldots,2^{nR_{2}}\}$, where $R\geq R_{1}+R_{2}$ and $I(U';Y,S)-3\epsilon > R$, such that to send message $M_{j},j\in \{2\ldots B\}$ given $M_{j}^{1}$ and $M_{j}^{2}$ transmit $X^{n}(M_{j}^{1}\oplus k^{s}_{j-1},M_{j}^{2}\oplus k^{f}_{j-1})\in C_{n}$ using Shannon's strategy \cite{C1} (one-time pad). The decoder uses joint typicality decoding together with the knowledge of the keys and the state information to decode message $\hat{M}_{j}$.
%%%%%%%%%%%%%%%%%%%%%%%%%%%%%%%%%%%%%%%%%%%%%%%%%%%%%%%%%%%%%%%%%%%%%%
%%%%%%%%%%%%%%%%%%%%%%%%%%%%%%%%%%%%%%%%%%%%%%%%%%%%%%%%%%%%%%%%%%%%%%
\off{
\subsection{Achievability When the Causal State Information is not Available at the encoder yet Available at the Decoder}
For the case where the state information is not available at the encoder yet available at the decoder, compared with the $\hat{R}_{1}$ case, we split the message $M_{j},j\in \{2,\ldots,B\}$, into two independent messages $M^{0}\in \{1,\ldots,2^{nR_{1}}\}$ and $M^{2}\in \{1,\ldots,2^{nR_{2}}\}$, where $R\geq R_{0} + R_{2}-3\epsilon$. To send message $M_{j},j\in \{2,\ldots,b\}$ given $M_{j}^{0}$ and $M_{j}^{2}$, transmit $X^{n}(M_{j}^{0},M_{j}^{2}\oplus k^{f}_{j-1})\in C_{n}$ using Shannon's strategy. The resulting rate is $R=I(U;Y|S)-I(U;Z)+R_{f}$.
}

\section{Upper Bound}\label{converse}
Assume $\epsilon_{n},\delta_{n}\rightarrow 0$ when $n\rightarrow \infty$. Consider first the two upper bounds, when the CSI is available at both transmitter and legitimate receiver, and a feedback at rate $R_f$ is present,
\begin{equation} \label{eq:R case 5 2}
R\leq \frac{1}{n}\sum_{i=1}^{n}I(X_{i};Y_{i}|S_{i})+\epsilon_{n}
\end{equation}
and
\begin{multline} \label{eq:R case 5 3}
R \leq  \frac{1}{n}\sum_{i=1}^{n}I(X_{i};Y_{i}|Z_{i},S_{i}) + \frac{1}{n}\sum_{i=1}^{n}H(S_{i}|Z_{i})\\
+ R_{f}+\delta_{n}.
\end{multline}

The upper bound in $(9)$ is since the secrecy capacity cannot be greater than the channel capacity.
Thus, \ref{converse theorem} follows by combining $(9)$ and $(10)$, then introducing a time sharing random variable \cite{C5} to show that the secrecy capacity must satisfy
\begin{multline} \label{eq:R case 5}
R  \leq  \max_{p(x|s)}\min\{I(X;Y|S),I(X;Y|Z,S)\\
 + H(S|Z)+R_{f}\}.
\end{multline}

We now continue similar to \cite{C8} using Fano's inequality, the fact that $L^{n}\rightarrow 0$ (the constraint on the secrecy) and the fact that $\frac{1}{n}\sum_{i=1}^{n}\log(|K_{i}^{f}|)\leq R_{f}$. Together with \ref{rec. lemma} below, a non-trivial extension of \cite[Lemma 1]{C8} for a wiretap with state information, the upper bound in $(10)$  will follow.

By Fano's inequality, for $\hat{M} = \hat{m}(Y^{n},K^{n}_{f},S^{n})$,
\begin{equation*}
H(M|\hat{M})\leq 1+P_{e}^{(n)}nR=n\epsilon_{n},
\end{equation*}
where $\epsilon_{n} \rightarrow 0$ as $n \rightarrow \infty $ if $P_{e}^{(n)}\rightarrow 0$. Since $\hat{M}$ is a function of $Y^{n},K^{n}_{f},S^{n}$,
\begin{eqnarray*}
H(M|Y^{n},K^{n}_{f},S^{n})&\leq& H(M|\hat{M})
\\
&\leq& n\epsilon_{n}.
\end{eqnarray*}
Using the secrecy constraint,
\begin{equation} \label{eq:R case 5 0}
I(M;Z^{n})=n\gamma_{n},
\end{equation}
where $\gamma_{n}\rightarrow 0$ as $n\rightarrow \infty$. Consequently,
\begin{eqnarray}
nR & = & H(M)\nonumber\\
& = & H(M|Z^{n})+I(M;Z^{n})\nonumber\\
& \stackrel{(a)}{=} & H(M|Z^{n})+n\gamma_{n}\nonumber\\
& = & I(M;Y^{n},K^{n}_{f},S^{n}|Z^{n})+H(M|Y^{n},Z^{n},S^{n},K^{n}_{f})\nonumber\\
&& + n\gamma_{n}\nonumber\\
& = & I(M;Y^{n},K^{n}_{f}|Z^{n},S^{n})+I(M;S^{n}|Z^{n})\nonumber\\
&& + H(M|Y^{n},Z^{n},S^{n},K^{n}_{f})+n\gamma_{n}\nonumber\\
& \stackrel{(b)}{\leq} & I(M;Y^{n},K^{n}_{f}|Z^{n},S^{n})+I(M;S^{n}|Z^{n})\nonumber\\
&& +n\epsilon_{n}+n\gamma_{n}\nonumber\\
& \stackrel{(c)}{=} & I(M;K^{n}_{f}|Z^{n},S^{n})+I(M;Y^{n}|K^{n}_{f},Z^{n},S^{n})\nonumber\\
&& +I(M;S^{n}|Z^{n})+n\delta_{n}\nonumber\\
& \stackrel{(d)}{\leq } & H(K^{n}_{f}|Z^{n},S^{n})+I(M,X^{n};Y^{n}|K^{n}_{f},Z^{n},S^{n})\nonumber\\
&& + H(S^{n}|Z^{n})+n\delta_{n}\nonumber
\end{eqnarray}
where (a) follows from $(12)$, (b) follows from Fano's inequality, and (c) follows by defining $\delta_{n}=\epsilon_{n}+\gamma_{n}$.
The following recursive lemma is now required.
\begin{lemma}\label{rec. lemma}
For each $j\in \{1,\ldots n\}$,
\small
\begin{multline*}
H(K^{j}_{f}|Z^{j},S^{j})+I(M,X^{j};Y^{j}|K^{j}_{f},Z^{j},S^{j})
 + H(S^{j}|Z^{j})
\\
 \leq  H(K^{j-1}_{f}|Z^{j-1},S^{j-1})
 + I(M,X^{j-1};Y^{j-1}|K^{j-1}_{f},Z^{j-1},S^{j-1})
\\
 +H(S^{j-1}|Z^{j-1})
+ H(K_{j}^{f}|M,X^{j-1},K^{j-1}_{f},Z^{j-1},S^{j-1})
\\
 + I(X_{j};Y_{j}|Z_{j},S_{j})+H(S_{j}|Z_{j}).
\end{multline*}
\normalsize
\end{lemma}
\begin{IEEEproof}
We start with the left hand side in the lemma, and show that one can indeed decrease $j$ to $j-1$ at the price of the added terms:
\small
\begin{eqnarray} \label{eq:R case 5 8}
&& \hspace{-1cm}H(K^{j}_{f}|Z^{j},S^{j})+I(M,X^{j};Y^{j}|K^{j}_{f},Z^{j},S^{j})
 +H(S^{j}|Z^{j})\nonumber\\
& = & H(K^{j}_{f}|Z^{j},S^{j})+I(M,X^{j};Y^{j}|K^{j}_{f},Z^{j},S^{j})\nonumber\\
&& + H(S^{j-1}|Z^{j})+H(S_{j}|Z^{j},S^{j-1})\nonumber\\
& \leq & H(K^{j}_{f}|Z^{j},S^{j})+I(M,X^{j};Y^{j}|K^{j}_{f},Z^{j},S^{j})\nonumber\\
&& +H(S^{j-1}|Z^{j-1})+H(S_{j}|Z_{j})\nonumber\\
& = & H(K^{j}_{f}|Z^{j},S^{j})+I(M,X^{j};Y^{j-1}|K^{j}_{f},Z^{j},S^{j})\nonumber\\
&& + I(M,X^{j};Y_{j}|Y^{j-1},K^{j}_{f},Z^{j},S^{j})\nonumber\\
&& +H(S^{j-1}|Z^{j-1})+H(S_{j}|Z_{j})\nonumber\\
& \leq & H(K^{j}_{f}|Z^{j},S^{j})+I(M,X^{j};Y^{j-1}|K^{j}_{f},Z^{j},S^{j})\nonumber\\
&& + I(M,Y^{j-1},K^{j}_{f},Z^{j-1},S^{j-1},X^{j};Y_{j}|Z_{j},S_{j})\nonumber\\
&& +H(S^{j-1}|Z^{j-1})+H(S_{j}|Z_{j})\nonumber\\
& \stackrel{(e)}{=} & H(K^{j}_{f}|Z^{j},S^{j})+I(M,X^{j};Y^{j-1}|K^{j}_{f},Z^{j},S^{j})\nonumber\\
&& +H(S^{j-1}|Z^{j-1})+I(X_{j};Y_{j}|Z_{j},S_{j})+H(S_{j}|Z_{j})\nonumber\\
& \leq & H(K^{j}_{f}|Z^{j},S^{j})\nonumber\\
&& +I(M,X^{j},Z_{j},S_{j};Y^{j-1}|K^{j}_{f},Z^{j-1},S^{j-1})\nonumber\\
&& +H(S^{j-1}|Z^{j-1})+I(X_{j};Y_{j}|Z_{j},S_{j})+H(S_{j}|Z_{j})\nonumber\\
& \stackrel{(f)}{=} & H(K^{j}_{f}|Z^{j},S^{j})
 +I(M,X^{j};Y^{j-1}|K^{j}_{f},Z^{j-1},S^{j-1})\nonumber\\
&& +H(S^{j-1}|Z^{j-1})+I(X_{j};Y_{j}|Z_{j},S_{j})+H(S_{j}|Z_{j})\nonumber\\
& = & H(K^{j}_{f}|Z^{j},S^{j})
 +I(M,X^{j-1};Y^{j-1}|K^{j}_{f},Z^{j-1},S^{j-1})\nonumber\\
&& +I(X_{j};Y^{j-1}|M,X^{j-1},K^{j}_{f},Z^{j-1},S^{j-1})\nonumber\\
&& +H(S^{j-1}|Z^{j-1})+I(X_{j};Y_{j}|Z_{j},S_{j})+H(S_{j}|Z_{j})\nonumber\\
& \stackrel{(g)}{=} & H(K^{j}_{f}|Z^{j},S^{j})
 +I(M,X^{j-1};Y^{j-1}|K^{j}_{f},Z^{j-1},S^{j-1})\nonumber\\
&& +H(S^{j-1}|Z^{j-1})+I(X_{j};Y_{j}|Z_{j},S_{j})+H(S_{j}|Z_{j})\nonumber\\
& = & H(K^{j}_{f}|Z^{j},S^{j})\nonumber\\
&& +I(M,X^{j-1},K_{j}^{f};Y^{j-1}|K^{j-1}_{f},Z^{j-1},S^{j-1})\nonumber\\
&& -I(K_{j}^{f};Y^{j-1}|K^{j-1}_{f},Z^{j-1},S^{j-1})\nonumber\\
&& +H(S^{j-1}|Z^{j-1})+I(X_{j};Y_{j}|Z_{j},S_{j})+H(S_{j}|Z_{j})\nonumber\\
& = & H(K^{j}_{f}|Z^{j},S^{j})\nonumber\\
&& +I(M,X^{j-1};Y^{j-1}|K^{j-1}_{f},Z^{j-1},S^{j-1})\nonumber\\
&& +I(K_{j}^{f};Y^{j-1}|M,X^{j-1},K^{j-1}_{f},Z^{j-1},S^{j-1})\nonumber\\
&& -I(K_{j}^{f};Y^{j-1}|K^{j-1}_{f},Z^{j-1},S^{j-1})\nonumber\\
&& +H(S^{j-1}|Z^{j-1})+I(X_{j};Y_{j}|Z_{j},S_{j})+H(S_{j}|Z_{j})\nonumber\\
& = & H(K^{j-1}_{f}|Z^{j},S^{j})+H(K_{j}^{f}|K^{j-1}_{f},Z^{j},S^{j})\nonumber\\
&& +I(M,X^{j-1};Y^{j-1}|K^{j-1}_{f},Z^{j-1},S^{j-1})\nonumber\\
&& +I(K_{j}^{f};Y^{j-1}|M,X^{j-1},K^{j-1}_{f},Z^{j-1},S^{j-1})\nonumber\\
&& +H(K_{j}^{f}|Y^{j-1},K^{j-1}_{f},Z^{j-1},S^{j-1})\nonumber\\
&& -H(K_{j}^{f}|K^{j-1}_{f},Z^{j-1},S^{j-1})\nonumber\\
&& +H(S^{j-1}|Z^{j-1})+I(X_{j};Y_{j}|Z_{j},S_{j})+H(S_{j}|Z_{j})\nonumber\\
& \stackrel{(h)}{\leq} & H(K^{j-1}_{f}|Z^{j},S^{j})\nonumber\\
&& +I(M,X^{j-1};Y^{j-1}|K^{j-1}_{f},Z^{j-1},S^{j-1})\nonumber\\
&& +I(K_{j}^{f};Y^{j-1}|M,X^{j-1},K^{j-1}_{f},Z^{j-1},S^{j-1})\nonumber\\
&& +H(K_{j}^{f}|Y^{j-1},K^{j-1}_{f},Z^{j-1},S^{j-1})\nonumber\\
&& +H(S^{j-1}|Z^{j-1})+I(X_{j};Y_{j}|Z_{j},S_{j})+H(S_{j}|Z_{j})\nonumber
\end{eqnarray}
\begin{eqnarray}
& = & H(K^{j-1}_{f}|Z^{j},S^{j})\nonumber\\
&& +I(M,X^{j-1};Y^{j-1}|K^{j-1}_{f},Z^{j-1},S^{j-1})\nonumber\\
&& +H(K_{j}^{f}|M,X^{j-1},K^{j-1}_{f},Z^{j-1},S^{j-1})\nonumber\\
&& -H(K_{j}^{f}|Y^{j-1},M,X^{j-1},K^{j-1}_{f},Z^{j-1},S^{j-1})\nonumber\\
&& +H(K_{j}^{f}|Y^{j-1},K^{j-1}_{f},Z^{j-1},S^{j-1})\nonumber\\
&& +H(S^{j-1}|Z^{j-1})+I(X_{j};Y_{j}|Z_{j},S_{j})+H(S_{j}|Z_{j})\nonumber\\
& \stackrel{(i)}{=} & H(K^{j-1}_{f}|Z^{j},S^{j})\nonumber\\
&& +I(M,X^{j-1};Y^{j-1}|K^{j-1}_{f},Z^{j-1},S^{j-1})\nonumber\\
&& +H(K_{j}^{f}|M,X^{j-1},K^{j-1}_{f},Z^{j-1},S^{j-1})\nonumber\\
&& +H(S^{j-1}|Z^{j-1})+I(X_{j};Y_{j}|Z_{j},S_{j})+H(S_{j}|Z_{j})\nonumber\\
& \stackrel{(j)}{\leq}& H(K^{j-1}_{f}|Z^{j-1},S^{j-1})\nonumber\\
&& +I(M,X^{j-1};Y^{j-1}|K^{j-1}_{f},Z^{j-1},S^{j-1})\nonumber\\
&& +H(K_{j}^{f}|M,X^{j-1},K^{j-1}_{f},Z^{j-1},S^{j-1})\nonumber\\
&& +H(S^{j-1}|Z^{j-1})+I(X_{j};Y_{j}|Z_{j},S_{j})+H(S_{j}|Z_{j})\nonumber
\end{eqnarray}
\normalsize
where (e) is due to the Markov chain $Y_{j}\leftrightarrow (X_{j},Z_{j},S_{j})\leftrightarrow (M,X^{j-1},K^{j}_{f},Y^{j-1},Z^{j-1},S^{j-1})$; (f) follows from $(S_{j},Z_{j})\leftrightarrow (M,X^{j},K^{j}_{f},Z^{j-1},S^{j-1})\leftrightarrow Y^{j-1}$. (g) is because $Y^{j-1}\leftrightarrow (M,X^{j-1},K^{j}_{f},Z^{j-1},S^{j-1})\leftrightarrow X_{j}$ form a Markov chain. (h) and (j) follow since conditioning reduces the entropy and (i) is due to the Markov chain $(M,X^{j-1})\leftrightarrow (Z^{j-1},Y^{j-1},K^{j-1}_{f},S^{j-1})\leftrightarrow K_{j}^{f}$.
\end{IEEEproof}
To continue, we use Lemma 1 recursively starting from (d):
\begin{eqnarray} \label{eq:R case 5 14}
nR & \leq & H(K^{n}_{f}|Z^{n},S^{n})+I(M,X^{n};Y^{n}|K^{n}_{f},Z^{n},S^{n})\nonumber\\
&& +H(S^{n}|Z^{n})+n\delta_{n}\nonumber\\
& \leq & H(K^{n-1}_{f}|Z^{n-1},S^{n-1})\nonumber\\
&& +I(M,X^{n-1};Y^{n-1}|K^{n-1}_{f},Z^{n-1},S^{n-1})\nonumber\\
&& +H(S^{n-1}|Z^{n-1})\nonumber\\
&& +I(X_{n};Y_{n}|Z_{n},S_{n})\nonumber\\
&& +H(S_{n}|Z_{n})+H(K_{n}^{f})+n\delta_{n}\nonumber\\
& \leq & H(K^{n-2}_{f}|Z^{n-2},S^{n-2})\nonumber\\
&& +I(M,X^{n-2};Y^{n-2}|K^{n-2}_{f},Z^{n-2},S^{n-2})\nonumber\\
&& +H(S^{n-2}|Z^{n-2})\nonumber\\
&& +I(X_{n-1};Y_{n-1}|Z_{n-1},S_{n-1})\nonumber\\
&& +I(S_{n-1}|Z_{n-1})+H(K_{n-1}^{f})\nonumber\\
&& +I(X_{n};Y_{n}|Z_{n},S_{n})\nonumber\\
&& +H(S_{n}|Z_{n})+H(K_{n}^{f})+n\delta_{n}\nonumber\\
& \leq & \ldots\nonumber\\
& \leq & \sum_{i=1}^{n}I(X_{i};Y_{i}|Z_{i},S_{i})+\sum_{i=1}^{n}H(S_{i}|Z_{i})\nonumber\\
&& +\sum_{i=1}^{n}H(K_{i}^{f})+n\delta_{n}.\nonumber
\end{eqnarray}
Thus,
\begin{eqnarray*} \label{eq:R case 5 15}
nR & \leq & \sum_{i=1}^{n}I(X_{i};Y_{i}|Z_{i},S_{i})+\sum_{i=1}^{n}H(S_{i}|Z_{i})\nonumber\\
&& +\sum_{i=1}^{n}H(K_{i}^{f})+n\delta_{n}.
\end{eqnarray*}
We now normalize by $n$, and use the constraint $\frac{1}{n}\sum_{i=1}^{n}\log(|K_{i}^{f}|)\leq R_{f}$. We have,
\begin{eqnarray*} \label{eq:R case 5 16}
R & \leq & \frac{1}{n}\sum_{i=1}^{n}I(X_{i};Y_{i}|Z_{i},S_{i})+\frac{1}{n}\sum_{i=1}^{n}H(S_{i}|Z_{i})+R_{f}+\delta_{n}.
\end{eqnarray*}
To conclude, the well-known technique of introducing a time sharing random variable is used. Assume $Q$ is independent of $X^{n},Y^{n},Z^{n},S^{n}$ and uniform on $\{1,\ldots,n\}$, this results in
%\small
\begin{eqnarray*}
R & \leq & R_{f}+\frac{1}{n}\sum_{i=1}^{n}(I(X_{i};Y_{i}|Z_{i},S_{i})+H(S_{i}|Z_{i}))+\delta_{n}\nonumber\\
%&& +H(S_{i}|Z_{i}))+\delta_{n}\nonumber\\
& = & R_{f}+\frac{1}{n}\sum_{i=1}^{n}(I(X_{i};Y_{i}|Z_{i},S_{i},Q=i)\nonumber\\
&& +H(S_{i}|Z_{i},Q=i))+\delta_{n}\nonumber\\
& = & R_{f}+I(X_{Q};Y_{Q}|Z_{Q},S_{Q},Q)+H(S_{Q}|Z_{Q},Q)+\delta_{n}\nonumber\\
%&& +H(S_{Q}|Z_{Q},Q)+\delta_{n}\nonumber\\
& = & R_{f}+I(X;Y|Z,S,Q)+H(S|Z,Q)+\delta_{n}%\nonumber\\
%&& +H(S|Z,Q)+\delta_{n}
\end{eqnarray*}
%\normalsize
where $X:=X_{Q}, Y:=Y_{Q}, Z:=Z_{Q}, S:=S_{Q}$.

Now, letting $n\rightarrow \infty$, we get $\delta_{n}\rightarrow 0$ and $\epsilon_{n}\rightarrow 0$, we have
\begin{eqnarray*} \label{eq:R case 5 20}
R & \leq & R_{f}+I(X;Y|Z,S,Q)+H(S|Z,Q)\nonumber\\
& \leq & R_{f}+I(X,Q;Y|Z,S)+H(S,Q|Z)\nonumber\\
& \leq & R_{f}+I(X;Y|Z,S)+H(S|Z).
\end{eqnarray*}
Similarly, it is also easy to see that
$R  \leq I(X;Y|S)$.\\
From the two above bounds,
\[
R\leq \min\{I(X;Y|S),I(X;Y|Z,S)+H(S|Z)+R_{f}\},
\]
and the theorem easily follows.

\section{Special Cases}\label{special}
We start with the simple reductions, which show that our result generalizes known results in the literature, and conclude with an interesting observation on the case where the CSI is given only to the legitimate receiver.
%%%%%%%%%%%%%%%%%%%%%%%%%%%%%%%%%%%%%%%%%%%%%%%%%%
\subsection{Degraded Channel with no Dependence on the State}
When $Z$ is a degraded\footnote{Note that \cite{C11} lists a few interesting cases for which its bounds are tight. We do not include this list here, and only referred to the degraded version.} version of $Y$ and $p(y,z|x,s)=p(y,z|x)$, we have
\begin{equation*}
C_{S} = \max_{p(x)}\min\{I(X;Y)-I(X;Z)+H(S)+R_{f},I(X;Y)\}.
\end{equation*}
That is, both $S$ and $R_f$ come into play as keys to increase the secrecy capacity of the \emph{canonical} DMWTC.
Of course, in the case there is no state information at all (yet the feedback is still present), the results coincide with those of \cite{C8} (Figure 6). That is, equation $(7)$ and the cases where it is tight.
\begin{figure}
  \centering
  \includegraphics[scale=1]{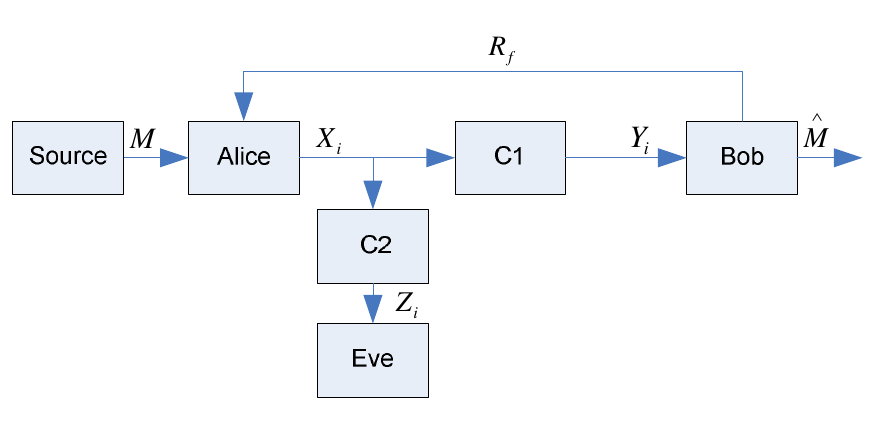}
  \caption{DMWTC in the presence of feedback.}
  \label{figure:6}
\end{figure}
%%%%%%%%%%%%%%%%%%%%%%%%%%%%%%%%%%%%%%%%
\subsection{Less Noisy Eavesdropper}
Consider the case where the output of the main channel is more noisy than the output of the eavesdropper channel and $p(y,z|x,s) = p(y,z|x)$. That is, $I(U;Z)\geq I(U;Y)$ for every $U$ such that $U\leftrightarrow X\leftrightarrow (Y,Z)$ from a Markov chain. We have
\[
I(U;Y|S)-I(U;Z|S)+H(S|Z)\leq H(S|Z)\leq H(S).
\]
Consequently, the secrecy capacity of this special class of channels is
\begin{equation*}
C_{S} = \max_{p(x)}\min\{H(S)+R_{f},I(X;Y)\}.
\end{equation*}
In this case, there is no benefit in a regular (Wyner-type) wiretap coding, and secrecy can be achieved \emph{only via the shared keys} (up to the main channel capacity).
%%%%%%%%%%%%%%%%%%%%%%%%%%%%%%%%%%%%%%%%%%%%%%%%%%
\subsection{No Feedback}
When the causal CSI is provide to the legitimate sender and legitimate receiver, yet the rate limited feedback is absent, the results easily reduce to those of Chia and El Gamal ~\cite{C11}. describes this case. The secrecy capacity is lower bounded by
\begin{multline}\label{eq:Special_Cases_1}
C_{S}  \geq \max\{\max_{p(u|s)p(x|u,s)}\min \{I(U;Y|S)-I(U;Z|S)\\
 + H(S|Z),I(U;Y|S)\},\\
 \max_{p(u)p(x|u,s)} \min\{H(S|Z,U),I(U;Y|S)\}\}.
\end{multline}
\begin{figure}
  \centering
  \includegraphics[scale=1]{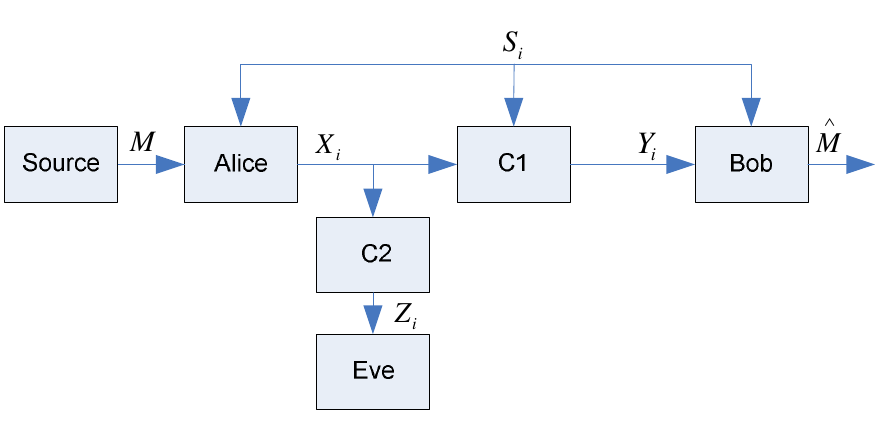}
  \caption{DMWTC with causal CSI.}
  \label{figure:7}
\end{figure}
%%%%%%%%%%%%%%%%%%%%%%%%%%%%%%%%%%%%%%%
\subsection{Causal State Information Only at the Legitimate Decoder}
Finally, we consider the following case where the rate limited feedback is available but the causal CSI is given only to the legitimate decoder. Figure 8 depicts this scenario. Note that \emph{without the feedback}, this case coincides with case 3 in ~\cite{C9}. However, even with feedback, when the side information is available only at the decoder, it can be viewed as part of the output, hence the results of \cite{C8} essentially apply with $Y'=(Y,S)$. Nevertheless, it is interesting to see that the same results can be achieved without sending keys from an outside source, and, rather, by feeding the state to the encoder. In particular, consider two possible achievable schemes. The first is similar to \cite{C8}, that is, use the feedback solely in order to send a key to the transmitter, and use this key to encrypt part of the message. The resulting lower bound is
\begin{eqnarray*}\label{eq:Special_Cases_2}
C_{S} & \geq & I(U;Y,S)-I(U;Z)+R_{f}\nonumber\\
& = & I(U;S)+I(U;Y|S)-I(U;Z)+R_{f}\nonumber\\
& = & I(U;Y|S)-I(U;Z)+R_{f},
\end{eqnarray*}
where, $S$ and $U$ are independent and the joint distribution is defined as $p(s)p(u)p(x|u)p(yz|x,s)$.
\begin{figure}
  \centering
  \includegraphics[scale=1]{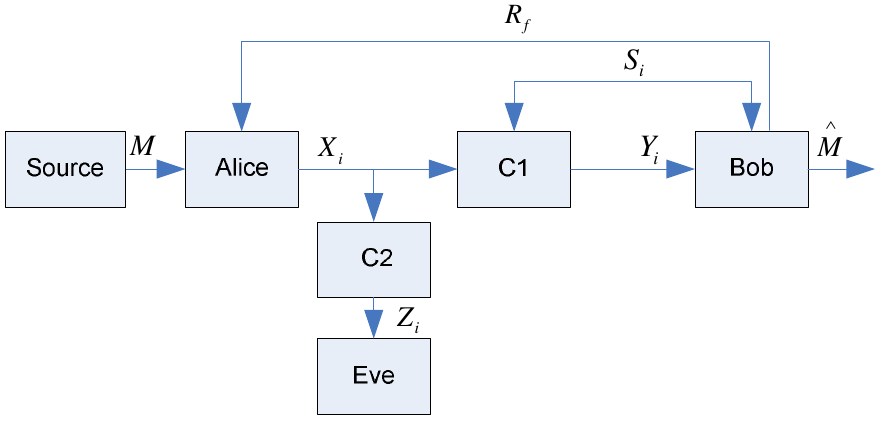}
  \caption{DMWTC with CSI only at the decoder and in the presence of rate limited feedback.}
  \label{figure:8}
\end{figure}

In the second, \emph{instead of sending fresh randomness through the feedback, the decoder sends the state sequence} (if the rate limit permits). Assume for now that $R_{f}=H(S)$. The encoder uses this information to \emph{both generate a key and optimize the main channel capacity}, as in the proof of the lower bound $\hat{R}_{1}$ when $R_{f}=H(S)$. In this case, we have
\begin{multline}\label{when sending state}
C_{S}  \geq  \max_{p(u|S)}\min \{I(U;Y|S)-I(U;Z|S)+H(S|Z),\\
I(U;Y|S)\}.
\end{multline}
We now show that the achievable scheme which results in $(13)$ is at least as good. We have
\begin{align*}
I&(U;Y|S)- I(U;Z)+H(S) \nonumber\\
 = & I(U;Y|S)-H(Z)+H(Z|U)+H(S)\nonumber\\
 \stackrel{(a)}{=} & I(U;Y|S)-H(Z)+H(Z|U,S)+H(S)\nonumber\\
 = & I(U;Y|S)-H(Z)+H(Z|U,S)+H(S,Z)-H(Z|S)\nonumber\\
 = & I(U;Y|S)-H(Z|S)+H(Z|U,S)+H(S|Z)\nonumber\\
 = & I(U;Y|S)-I(U;Z|S)+H(S|Z),
\end{align*}
where (a) follows from the Markov chain $S\leftrightarrow U\leftrightarrow Z$.
While the two expressions are equal, it is clear that the later achievability scheme is at least as good as the optimization is over all possible $p(u|s)$ and not simply $p(u)$. Of course, since \cite{C8} applies here, the benefit can only be in avoiding the need for an outside source of randomness.

When $R_{f}>H(S)$, the same scheme can be used, yet, in addition to sending the state information from the decoder to the encoder through the secure rate limited feedback, one can send also a key $K_f$, with $H(K_f)<R_{f}-H(S)$. The encoding procedure is as in the proof of the lower bound $\hat{R}_{1}$. For this case, the resulting secrecy capacity is bounded by
\begin{multline*}
C_{S}  \geq  \max_{p(u|s)}\min \{I(U;Y|S)-I(U;Z|S) \\+H(S|Z)+H(K_f),I(U;Y|S)\}.
\end{multline*}
Note that, as inferred from \cite{ahlswede2006transmission} and re-assured in \cite{C8}, the best use of the feedback is in sending  a random key, hence, there is no need to send $Y$ or a compressed version of it in the extra rate above $H(S)$.

When $R_{f}<H(S)$, we conjecture that the preferred scheme is to feedback the \emph{compressed state information} $S'$, and use it to extract \emph{common randomness}, though, of course, at rate smaller than $H(S)$.

Finally, note that in the case where there is state information only at the encoder and, in addition, a secure feedback is available (Figure 8 ), similar arguments to the one used above can be used. Thus, we conjecture that by case 4 in ~\cite{C9}, ~\cite{C11}, and when adding the secure rate limited feedback, the resulting bound is
\begin{multline*}
C_{S} \geq \max_{p(u,x|s)} \min \{I(U;Y)-I(U;Z|S)+H(S|Z)+R_{f},\\ I(U;Y)-I(U;S)\}.
\end{multline*}
\begin{figure}
  \centering
  \includegraphics[scale=1]{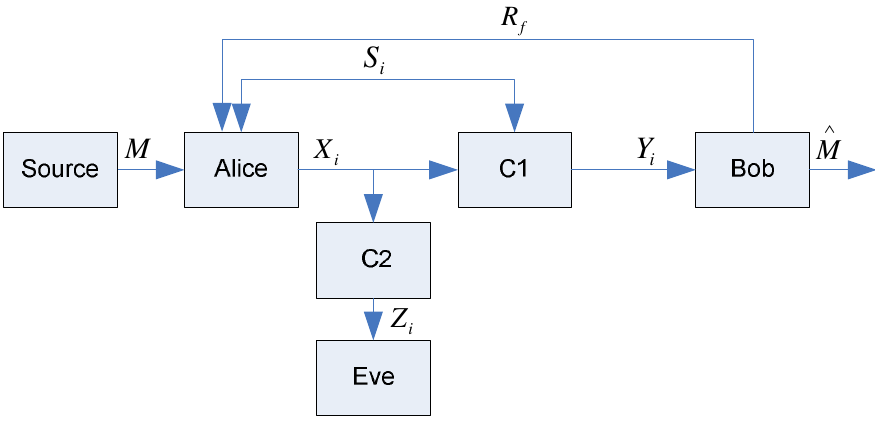}
  \caption{Wiretap Channel With State Information In The Encoder And Rate Limited Feedback.}
  \label{figure:9}
\end{figure}

\section{Conclusions}\label{conc}
Physical layer security promises to achieve secret transmission at the expense of transmission rate. While several models in this area, such as the wiretap channel, are well understood, with both capacity results and practical codes, more complex scenarios are still unsolved. For example, in order to apply the concepts of physical layer security to networks with state information and two way communication, the canonical model of a wiretap channel with state and feedback should be understood.

In this paper, the wiretap channel with  causal state information and secure rate limited feedback at the encoder and legitimate decoder is studied. We established upper and lower bounds on the secrecy capacity, and proved their tightness in the case of a less capable eavesdropper. The suggested coding scheme is based on two steps of key generation, one from the causal state information and one from fresh randomness through the rate limited feedback. It was shown that in several special cases, the results reduce to known expressions in the literature.

\section*{Acknowledgment}
This research was partially supported by DSP Group inc.

\bibliographystyle{IEEE}
\bibliography{wiretap}
\end{document}